\documentclass[a4paper]{ifacconf}

\usepackage{graphicx,amsmath,url}      
\usepackage[round]{natbib}             

\usepackage[english]{babel}

\usepackage[T1]{fontenc}
\usepackage[utf8]{inputenc}
\usepackage{ae}

\begin{document}


\begin{frontmatter}

\title{HR-Agents: Using Multiple LLM-based Agents to Improve Q\&A about Brazilian Labor Legislation} 


\author{Gabriel K. Moraes},
\author{Gabriel S. M. Dias},
\author{Vitor L. Fabris},
\author{Lucas D. Gessoni},
\author{Leonardo R. do Nascimento},
\author{Charles S. Oliveira},
\author{Vitor G. C. B. de Farias},
\author{Fabiana C. Q. de O. Marucci},
\author{Matheus H. R. Vicente},
\author{Gabriel U. Talasso},
\author{Erik Soares},
\author{Amparo Munoz},
\author{Sildolfo Gomes},
\author{Maria L. A. de S. Cruvinel},
\author{Leonardo T. dos Santos},
\author{Renata De Paris} and
\author{Wandemberg S. P. Gibaut}

\address{Eldorado Institute of Research – Av. Alan Turing, 275, Cidade Universitária, 13.083-898 – Campinas – SP – Brazil \\(e-mail: \{gabriel.moraes, gabriel.martins, vitor.fabris, lucas.gessoni, leonardo.nascimento, charles.oliveira, vitor.farias, fabiana.marucci, matheus.rosado, gabriel.talasso, erik.soares,   amparo.munoz, sildolfo.gomes, maria.cruvinel, leonardo.santos, renata.paris, wandemberg.gibaut,\}@eldorado.org.br)}



\renewcommand{\abstractname}{{\bf Abstract:}}

\begin{abstract}
The Consolidation of Labor Laws (CLT) serves as the primary legal framework governing labor relations in Brazil, ensuring essential protections for workers. However, its complexity creates challenges for Human Resources (HR) professionals in navigating regulations and ensuring compliance. Traditional methods for addressing labor law inquiries often lead to inefficiencies, delays, and inconsistencies. To enhance the accuracy and efficiency of legal question-answering (Q\&A), a multi-agent system powered by Large Language Models (LLMs) is introduced. This approach employs specialized agents to address distinct aspects of employment law while integrating Retrieval-Augmented Generation (RAG) to enhance contextual relevance. Implemented using CrewAI, the system enables cooperative agent interactions, ensuring response validation and reducing misinformation. The effectiveness of this framework is evaluated through a comparison with a baseline RAG pipeline utilizing a single LLM, using automated metrics such as BLEU, LLM-as-judge evaluations, and expert human assessments. Results indicate that the multi-agent approach improves response coherence and correctness, providing a more reliable and efficient solution for HR professionals. This study contributes to AI-driven legal assistance by demonstrating the potential of multi-agent LLM architectures in improving labor law compliance and streamlining HR operations.
\end{abstract}


\end{frontmatter}


\section{Introduction}
\label{s.introduction}

The Consolidation of Labor Laws (CLT) \citep{BRASIL1943} is the main legal framework governing labor relations in Brazil, ensuring key rights such as working hour regulations, minimum wage, vacation entitlements, and protections against unjust dismissal. Influenced by both civil and common law traditions, the CLT follows a corporatist model, where state intervention mediates labor relations to promote social justice \citep{cabral_direito}. However, over the decades, this model has faced criticism for its rigidity and the challenges it poses in adapting to the dynamic demands of a globalized economy \citep{silva_justica}.



Given the complexity and extensive nature of CLT regulations, Human Resources (HR) departments frequently encounter intricate and detailed legal inquiries related to employment laws, compliance, and workplace policies. Recently, Large Language Models (LLMs) have emerged as powerful tools for enhancing the efficiency and accuracy of information retrieval and response generation across diverse domains, streamlining complex inquiries and reducing operational burdens \citep{zhao2023survey}. Furthermore, previous research on multi-agent systems has demonstrated additional potential in verifying responses, reducing misinformation, and improving accuracy \citep{wu2023autogen, talebirad2023multi}, as well as enhancing efficiency in obtaining legal information \citep{blair2023can, minaee2024large}. Nevertheless, to the best of our knowledge, no prior work has specifically implemented a multi-agent LLM tailored explicitly for the HR legal domain.


This study addresses this gap by introducing a specialized multi-agent LLM system explicitly designed to support HR professionals in navigating CLT labor legislation. The proposed framework leverages collaborative interactions among multiple specialized agents and employs Retrieval-Augmented Generation (RAG) methodologies \citep{gao2023retrieval} to generate detailed, contextually appropriate, and legally accurate responses. This multi-agent approach addresses an existing research gap by tailoring advanced language modeling techniques specifically to the complexities of the HR legal domain.

The main contributions of this work are:
\begin{itemize}
    \item The presentation of an multi-agent LLM-based Q\&A solution aligned with Human Preferences to answer questions about Brazilian labor law;
    \item A study evaluating the effectiveness of a multi-agent LLM framework in the HR legal domain, comparing it to a RAG pipeline with a single LLM providing CLT-contextualized responses.\end{itemize}

\section{Theoretical background}
\label{s.theoretical}

\subsection{Large Language Models}
\label{ss.th.llm}

LLMs are advanced language models based on the Transformer architecture \citep{vaswani2017attention}, designed to process and generate human language with remarkable flexibility and capability. By leveraging massive parameter sizes, extensive datasets, and significant computational power, LLMs represent the latest generation in the evolution of language models, profoundly transforming Natural Language Processing (NLP) \citep{zhao2023survey}. At their core, Transformers rely on the self-attention mechanism \citep{vaswani2017attention}, which enables efficient processing of sequential data by capturing long-range dependencies while allowing parallelization during training. These characteristics make LLMs highly scalable, capable of handling vast datasets, and adaptable to diverse linguistic contexts.

LLMs can scale to hundreds of billions of parameters or more, and they demonstrate remarkable emergent capabilities. These include: In-Context Learning \citep{brown2020language}, where models learn new tasks from a few examples provided at inference time without additional training; Instruction Following \citep{chung2024scaling}, where they generalize to new tasks based solely on instructions—even in the absence of examples—after fine-tuning; and Multi-Step Reasoning, where they decompose complex problems into intermediate steps using techniques such as chain-of-thought prompting \citep{wei2022chain}.

\subsection{LLM-based Agents}
\label{ss.th.llm-agents}


LLM-based agents are designed to process information, make decisions, and perform complex tasks by utilizing resources such as Short-Term and Long-Term Memory. They rely on historical data to contextualize and retain relevant information throughout their operation. Moreover, these agents can interface with external tools, including APIs, scripts, and web browsers, facilitating capabilities like web scraping. This integration not only enhances the agents' ability to retrieve and process real-time data but also optimizes their performance and accuracy in multi-agent systems. As a result, this approach significantly augments the agents' potential for collaborative problem-solving, decision-making, and adaptability across diverse operational contexts \citep{zhang2024llm, shinn2024reflexion}.

CrewAI is a framework that allows the creation of environments with multiple agents, each assigned distinct roles, background stories and specifics descriptions. This structure is essential for ensuring that agents can validate outputs before advancing to the next stage. Additionally, CrewAI allows for individual agent configuration, including the selection of which LLM each agent will use, and the specification of attributes such as task delegation to designated agents and the integration of custom tools for task execution.

The CrewAI framework supports the use of various tools, including RAG. To improve development efficiency and modularity, specific code snippets can be created to configure the parameters required for implementing RAG within the context of a specific use case. This approach enables a more structured and organized integration of RAG resources, ensuring alignment with the project's specific requirements.




\section{Methodology}
\label{s.methodology}

\subsection{LLM Preliminary Accuracy Comparison}
\label{ss.me.comparison}

In this study, we conducted an analysis involving various models, including Llama 2 (7B), Llama 3 (8B), and Llama 3.1 (8B), along with models based from Llama's architecture, such as Bode 7B \citep{bode7b}, Sabiá 2 \citep{sabia2}, and Periquito 3B \citep{periquito3b}  which was derived from a smaller model (Open-Llama 3B). Additionally, OpenAI models - GPT 3.5, GPT-4o, and GPT-4o-mini - were also included in the evaluation.

The selection of LLMs was guided by responses to 19 questions focused on Human Resources and Brazilian labor laws. These questions were developed by legal experts to reflect common concerns raised by workers. The ground truth for the responses was established using the specialists’ expertise and references from the CLT, ensuring a reliable basis for evaluation. Following initial testing, the models were ranked according to the quality of their answers, considering aspects such as coherence, clarity, hallucination frequency, and code-switching. Based on this assessment, GPT 4o, and Llama 3.1 8B \citep{meta2024llama3.1} were selected for their precision and consistent accuracy in delivering trustworthy responses.




\subsection{HR-Agents System}
\label{ss.me.hragents}

For this project, the CrewAI framework was used to create multiple agents, one of which has access to an external custom RAG tool. A structure comprising three agents was defined, with the aim of promoting cooperation among them by passing input sequentially along the workflow. The roles of each agent are as follows:

\begin{itemize}

    \item Office Clerk: This agent is responsible for filtering the relevant information from the user's question and forwarding it to the appropriate output. If the question is unrelated to the Human Resources context, the agent will direct the inquiry to another processing flow.

    \item HR Specialist: This agent’s role is to filter the information contained in the PDF dataset and, based on the previously described RAG (retrieval-augmented generation), formulate appropriate responses.

    \item Chief of Human Resource Department: The main role of this agent is to review the responses generated by the previous agents, with the goal of guaranteeing the clarity and coherence of the information provided. If necessary, the agent makes corrections to ensure that the responses are well-structured and consistent.

\end{itemize}

To process the document with precision, the tool uses a Character Text Splitter which breaks the text into smaller, manageable chunks. These chunks are structured to include overlapping sections, ensuring that the ideas flow continuously between them. By setting a chunk size of 512 characters and an overlap of 256, the tool reaches a balance between comprehensiveness and usability, making it easier for language models to process and understand the dataset's data.


RAG was employed to provide relevant context from the CLT to the LLMs for answering questions. The source CLT document was segmented into individual articles using the delimiter “Art.”, with each segment representing a distinct article. This process produced a structured collection of textual chunks, each corresponding to a logically coherent and independently retrievable section of the original text.

Following segmentation, dense embedding vectors were generated using the Multilingual E5 text embedding model \citep{wang2024multilingual}. Retrieval was performed using the FAISS library \citep{douze2024faiss}, configured to identify the top 20 most semantically similar document segments based on the cosine similarity between the query vector and document embeddings. The retrieved document segments were then concatenated into a single textual context provided to the LLM.

\subsection{Evaluation Metrics}
\label{ss.me.metrics}

Using the same set of questions from the initial model selection test, the evaluation process in this study employed four distinct metrics: one statistical (BLEU), two based on the "LLM-as-judge" approach—answer similarity and answer correctness—and one based on subjective human evaluation.

\subsubsection{BLEU}

The BiLingual Evaluation Understudy (BLEU) \citep{papineni2002bleu} is a metric  initialy for evaluate machine translations to measure the quality of text generated by machine translation systems or language generation models.
Comparing the generated text with one or more references texts, it calculate the precision of n-grams (sequences of n words) that match between the generated text and the references, making BLEU an effective choice for tasks where exact wording is important. However, BLEU has limitations when evaluating tasks that allow semantic variation, as it does not account for synonyms or paraphrases.

\subsubsection{Answer Similarity}

Using LLM itself as a "judge" is a one approach to evaluate complex answers from others LLM's, focusing on how well they align with reference answers in terms of content and relevance.

Answer Similarity measures how closely the generated answer matches the reference answer, which is often provided by human evaluators. It is essential in tasks such as question answering and, where exists multiple valid answers. The goal is not for the answers to be identical but for them to convey the same meaning and content. Metrics like cosine similarity, Jaccard similarity, and BLEU score are commonly used to quantify answer similarity.

In this project, the Cosine Similarity metric was utilized through the RAGAS framework, which allows for a more context-aware evaluation of generated answers. By leveraging vectorized text representations, Cosine Similarity measures the angular distance between the generated and reference answers, ensuring that the assessment captures semantic relationships beyond simple lexical overlap.

\subsubsection{Answer Correctness}

Answer Correctness is one type of evaluation where measures how accurately a model's generated response reflects the factual content and semantic meaning of the ground truth answer. It is computed by considering both factual correctness and semantic similarity. 
Factual correctness measures the overlap between the generated answer and the ground truth in terms of true positive (TP), false positive (FP), and false negative (FN) facts. TP represents facts that are present in both the generated answer and the reference answer, while FP and FN refer to discrepancies between the two. Additionally, semantic similarity evaluates how closely the meaning of the generated answer aligns with the reference answer. These combined factors help assess the overall correctness of the model's response.

\subsubsection{Subjective Evaluation Metrics}

Recognizing the limitations of automated lexical metrics like BLEU, we conducted complementary subjective evaluations to better assess LLM performance. These focused on two key qualitative dimensions: response precision and response quality.

LLM responses to 19 expert-formulated questions in labor law and HR were systematically evaluated. Response precision measured the legal accuracy and contextual appropriateness of the answers. Response quality assessed coherence, clarity, relevance, and logical flow.

Labor law and HR specialists performed these evaluations manually, capturing critical aspects often missed by automated metrics, such as interpretative accuracy, tone, regulatory compliance, and implicit meaning. These expert insights provided essential human-centered validation, complementing objective metrics and ensuring a more reliable assessment of LLM performance in real-world legal contexts.


\section{Results}
\label{s.results}

After collecting responses and metrics, the results were obtained and the analysis assessed the performance of the models comparing RAG and CrewAI Agent responses, providing insights into the effectiveness of the LLMs in generating answers.

Figures 1 and 2 present a BLEU score comparison between GPT-4o and Llama 3.1 8B, showing improved results for the GPT-4o when using CrewAI agents. This improvement may be attributed to the nature of GPT's responses, which were generally brief and often replying with statements like “I don't know”. In contrast, Llama produced longer and more detailed answers, though these occasionally introduced discrepancies or inaccuracies. The higher BLUE scores for GTP-4o might be linked to its shorter responses, which resulted in a higher overlap of words with the ground truth.

It's also important to note that the BLEU metric evaluates responses by comparing their n-grams to those in the ground truth. Given the complexity of certain questions, the wording in the LLMs' responses may differ from the reference answers, even if the meaning is accurate. This lexical variation can result in lower BLEU scores despite the presence of valid content.

\begin{figure}[!htbp]
  \centering
    \includegraphics[width=0.45\textwidth]{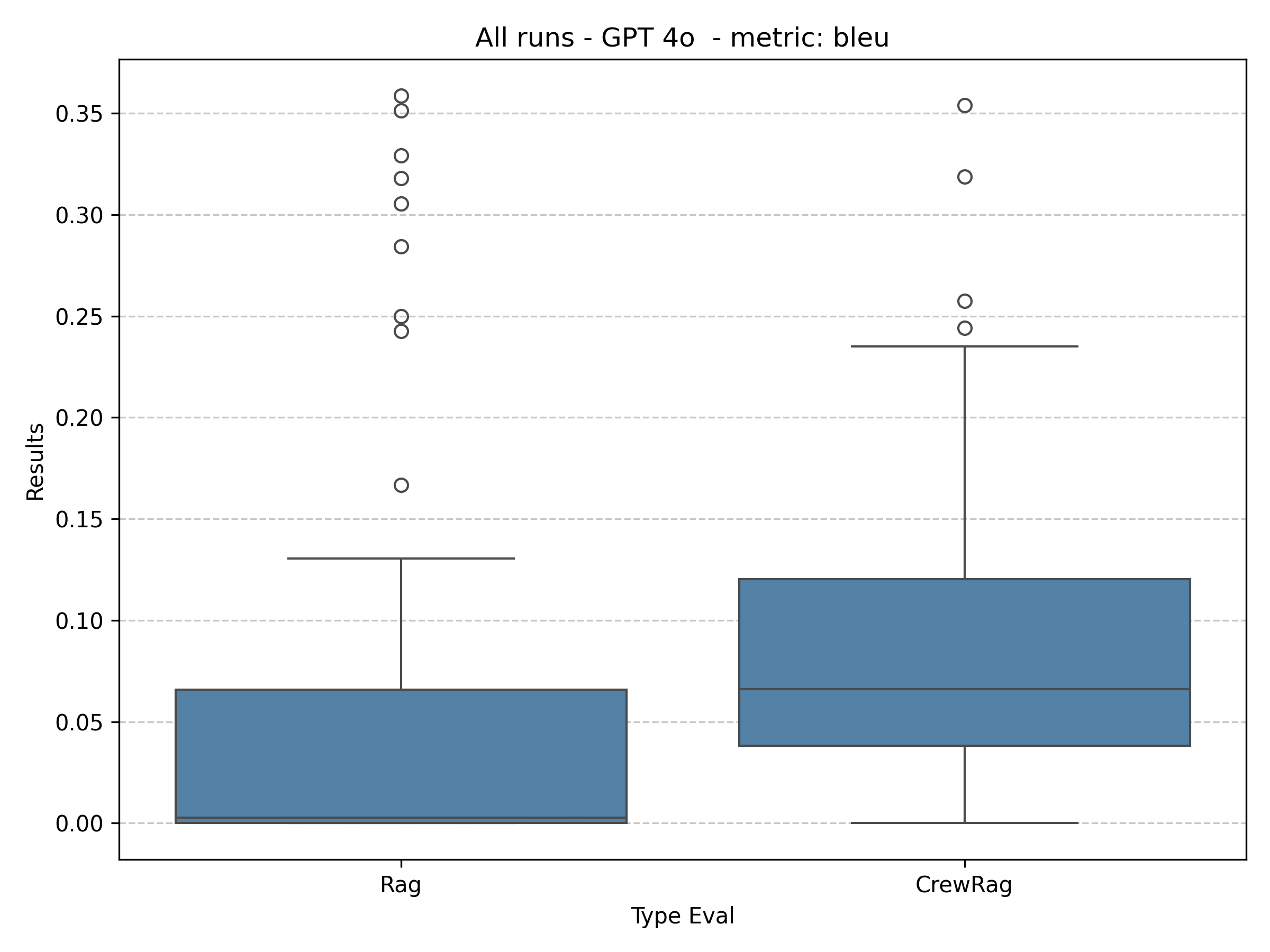}
  \caption{GPT 4o BLEU evaluation}
  \label{fig:GPT_4o_All_runs_metric_bleu}
\end{figure}

\begin{figure}[!htbp]
  \centering
    \includegraphics[width=0.45\textwidth]{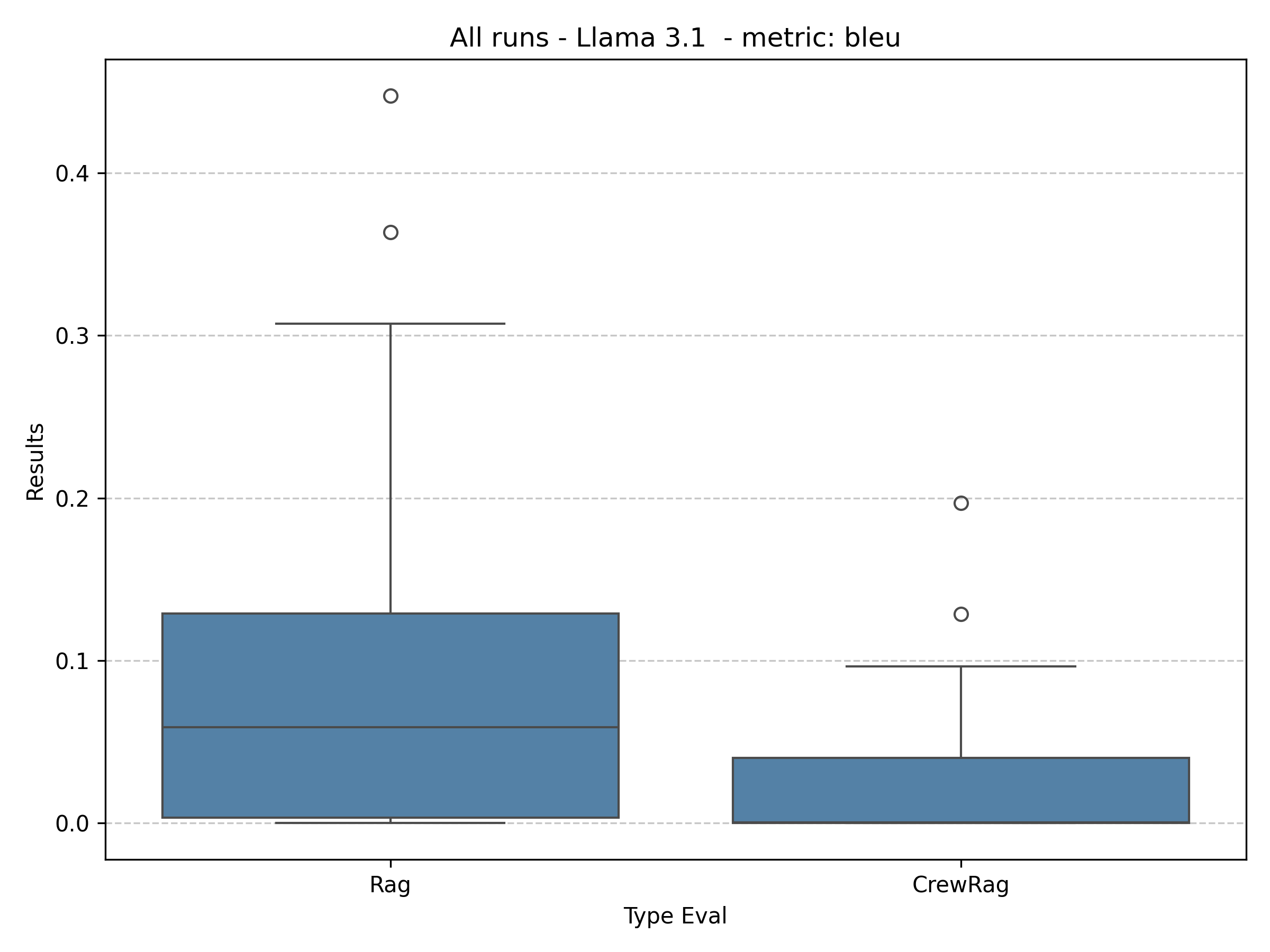}
  \caption{Llama 3.1 8B BLEU evaluation}
  \label{fig:images/LLama_All_runs_metric_rougeL.png}
\end{figure}




Figures 3 and 4 display the answer similarity scores, indicating that GPT-4o outperformed Llama within the CrewAI agent configuration, although both models demonstrated satisfactory lexical alignment with the ground truth across both RAG and CrewAI settings. These results indicate that, in the tests conducted, both models were generally capable of producing responses containing terms closely aligned with those found in the reference answers.

However, it is important to note that the answer similarity metric evaluates textual overlap and does not penalize semantically similar responses that may still arrive at factually incorrect conclusions. To address this limitation, the answer correctness metric is employed, offering a more rigorous assessment of the factual alignment between model outputs and the ground truth.

\begin{figure}[!htbp]
	\centering
	\includegraphics[width=0.45\textwidth]{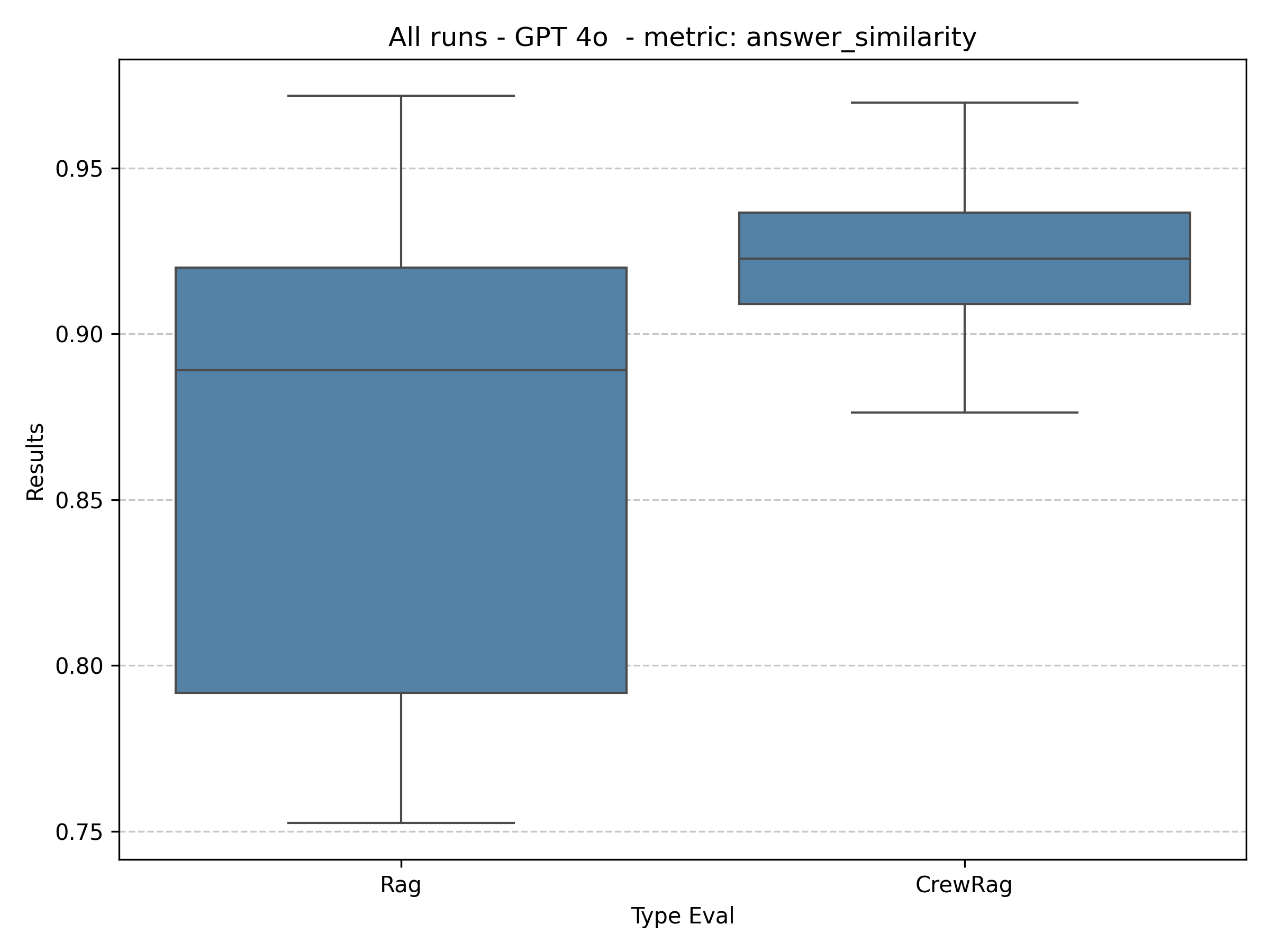}
	\caption{GPT 4o Answer Similarity }
	\label{fig:GPT_4o_All_runs_metric_answer_similarity}
\end{figure}

\begin{figure}[!htbp]
	\centering
	\includegraphics[width=0.45\textwidth]{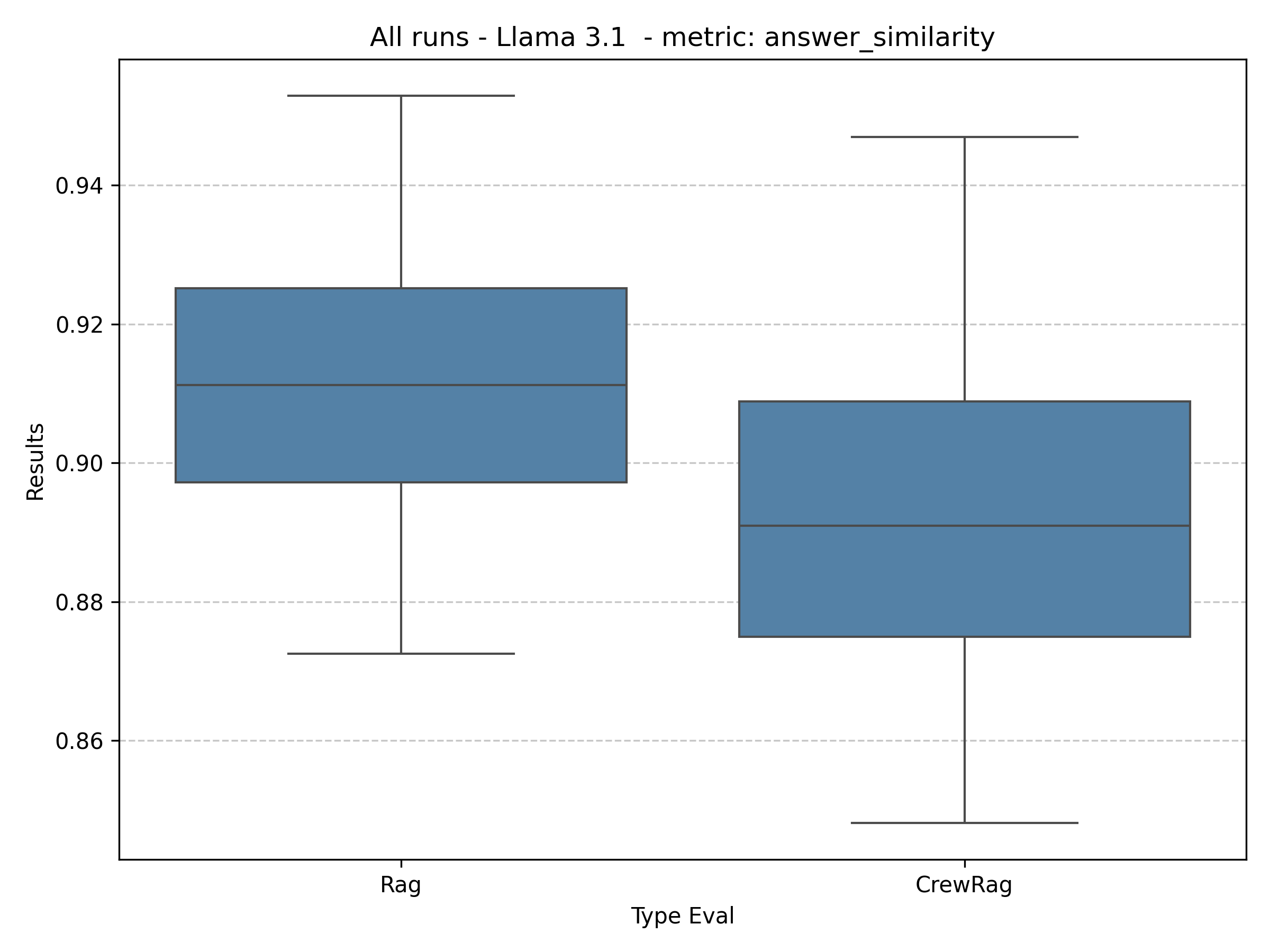}
	\caption{Llama 3.1 8B Answer Similarity }
	\label{fig:LLama_All_runs_metric_answer_similarity}
\end{figure}

Based on the results showed in Figures 5 and 6, RAG consistently achieved higher answer correctness scores for both GPT-4o and Llama 3.1 8B, with median values of 0.67 and 0.60, respectively. These results highlight RAG’s effectiveness in grounding answers with retrieved legal content, which enhances factual accuracy and semantic alignment with the ground truth. While the advantage was clear, the differences were not substantial when compared to CrewAI agents, which scored 0.55 for GPT-4o and 0.54 for Llama—demonstrating that agent-based cooperation can approximate similar levels of correctness even without explicit retrieval.

These findings suggest that while RAG improves factual grounding, CrewAI agents offer a competitive alternative by leveraging role-based task handling and multi-agent reasoning. The relatively small performance gap underscores the potential for further enhancing agent coordination or integrating fine-tuned legal modules.


\begin{figure}[!htbp]
  \centering
    \includegraphics[width=0.45\textwidth]{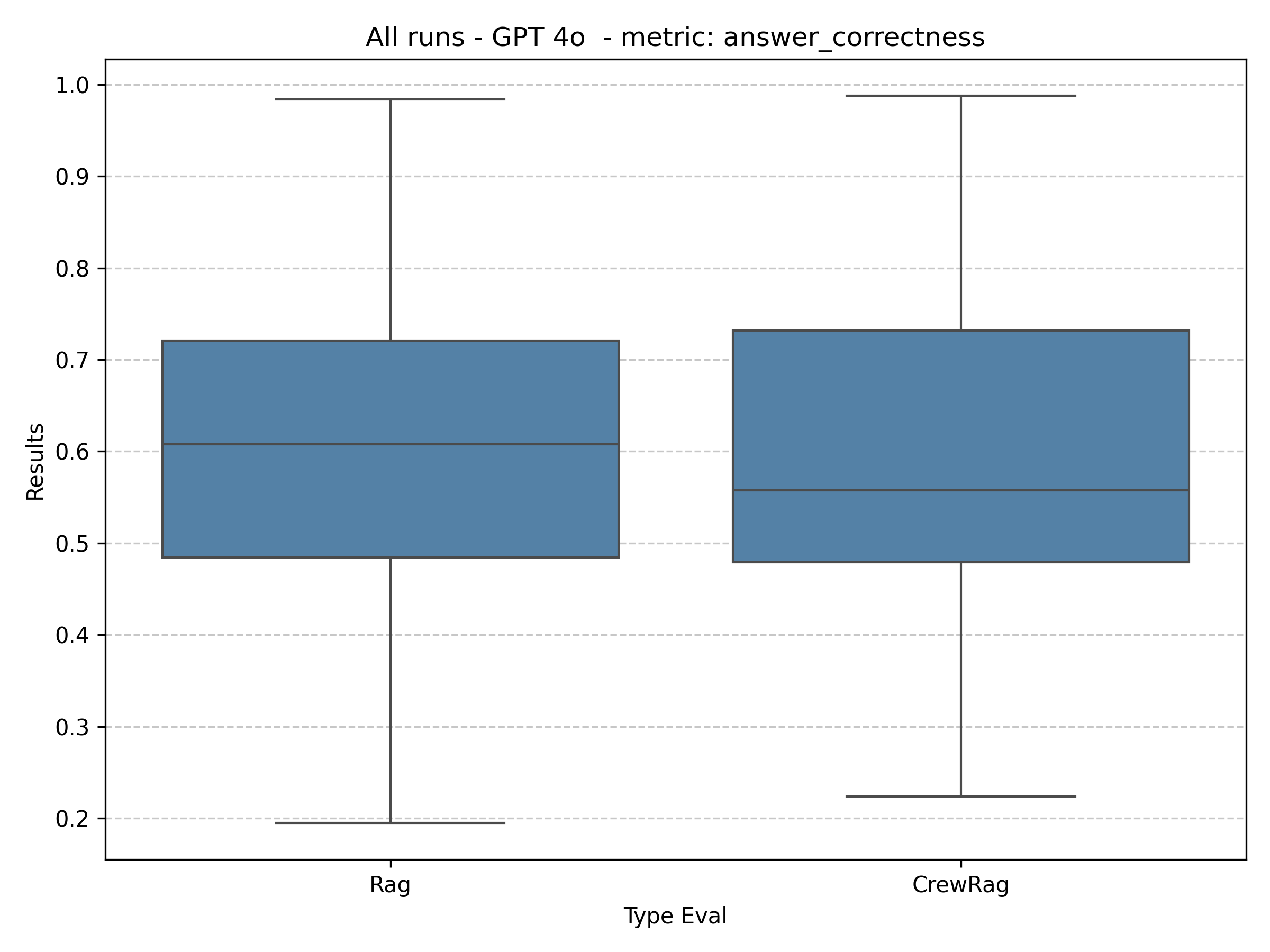}
  \caption{GPT 4o Answer Correctness results }
  \label{fig:GPT_4o_All_runs_metric_answer_correctness}
\end{figure}

\begin{figure}[!htbp]
  \centering
    \includegraphics[width=0.45\textwidth]{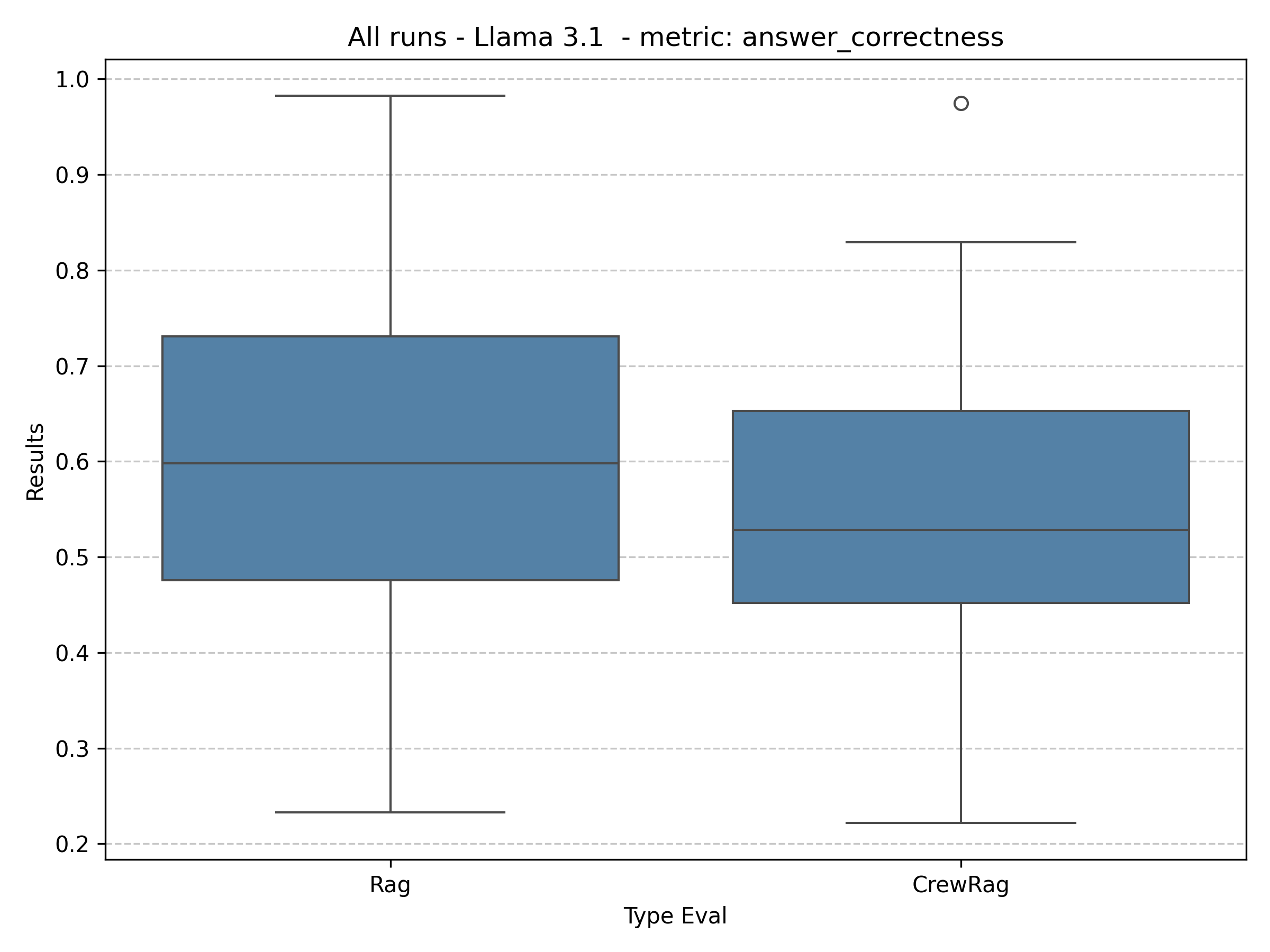}
  \caption{Llama 3.1 8B Answer Correctness results }
  \label{fig:LLama_All_runs_metric_answer_correctness}
\end{figure}

\begin{table}[ht]
    \centering
    \begin{tabular}{|l|c|c|}
    \hline
    \textbf{} & \textbf{Precision} & \textbf{Quality} \\
    \hline
    mean     & 6.47   & 9.75 \\
    median   & 8.00   & 10.00 \\
    std      & 3.43   & 0.78 \\
    var      & 11.79  & 0.61 \\
    min      & 0.00   & 6.00 \\
    max      & 10.00  & 10.00 \\
    range    & 10.00  & 4.00 \\
    cv (\%)  & 53.06  & 8.05 \\
    \hline
    \end{tabular}
    \caption{GPT 4o - Subjective evaluation}
    \label{tab:estatisticas}
    \end{table}
    
Table 1 reports the results of the subjective evaluation for GPT-4o, revealing a mean precision score of 6.47 with substantial variability, suggesting inconsistency in the model’s ability to provide accurate responses. In contrast, the model demonstrates a high mean quality score of 9.75, indicating that despite occasional imprecision, the responses were consistently well-articulated and coherent. These results underscore a key area for improvement in precision while reinforcing the model’s strength in generating high-quality outputs.

\begin{table}[ht]
\centering
\begin{tabular}{|l|c|c|}
\hline
\textbf{} & \textbf{Precision} & \textbf{Quality} \\
\hline
mean     & 4.72  & 8.72  \\
median   & 5.00  & 9.00  \\
std      & 3.44  & 1.34  \\
var      & 11.85 & 1.80  \\
min      & 1.00  & 1.00  \\
max      & 10.00 & 10.00 \\
range    & 9.00  & 9.00  \\
cv (\%)  & 72.94 & 15.38 \\
\hline
\end{tabular}
\caption{Llama 3.1 8B - Subjective evaluation}
\label{tab:estatisticas}
\end{table}

Table 2 presents the subjective evaluation statistics for Llama 3.1 8B. The results indicate that precision demonstrates considerable variability, with a mean score of 4.72 and a coefficient of variation of 72.94\%, suggesting notable inconsistencies in evaluators' assessments. In contrast, response quality appears more stable, with a mean score of 8.72 and a coefficient of variation of 15.38\%, reflecting greater consistency across evaluations. When compared to the earlier findings for GPT-4o, precision remains a key area of concern—exhibiting even higher variability—whereas quality continues to be relatively high and stable, albeit marginally lower than GPT-4o’s mean quality score of 9.75.


It is important to acknowledge that while automated metrics such as BLEU offer valuable insights into model performance, these metrics may not fully capture the intricacies and diverse interpretations inherent to the CLT's legal definitions. Recognizing this limitation, we complemented our evaluation approach with detailed human assessments conducted by HR and legal specialists, ensuring that critical nuances, contextual subtleties, and interpretative variations specific to labor law were appropriately considered. This combined methodology provides a more holistic and robust evaluation of our proposed system, underlining the essential role of human judgment alongside automated measures in complex legal domains.


\section{Conclusion}
\label{s.conclusion}

The creation and integration of environments containing agents with LLMs through frameworks such as CrewAI have facilitated the understanding of this technology, while also enabling the use of customized tools tailored to specific purposes. Applied to the context of Human Resources and labor laws, this study highlights, through various comparisons and evaluation methods, the impact of using such technology in specific contexts and the challenges it entails.

While subtle improvements are noticeable in certain metrics, a meticulous analysis of each generated response, combined with subjective evaluation, revealed that the model was capable of producing responses consistent with expectations. It was able to reference relevant laws and articles related to each question, making it feasible for implementation in chatbot systems focused on Brazilian labor laws.

Looking ahead, this study lays a solid foundation for expanding the application of multi-agent LLM systems to legal domains beyond the CLT framework. The proposed architecture demonstrates a high degree of adaptability, enabling the integration of alternative LLM models and the refinement of existing agent configurations through tailored enhancements. Future research can explore the use of domain-specific language models trained on targeted legal corpora, integrating them into the multi-agent pipeline to improve accuracy and contextual understanding in specialized areas of law.

Moreover, upcoming investigations could prioritize the development of evaluation metrics tailored to legal reasoning. Traditional metrics, such as BLEU, may not fully capture the depth and nuance required for legal assessments. Addressing these limitations—alongside inconsistencies like GPT-4o’s overly concise answers or Llama’s tendency toward verbose but occasionally imprecise outputs—will be essential for fine-tuning model behavior. Collectively, these research directions can drive significant improvements in the performance, flexibility, and real-world applicability of multi-agent LLM systems across complex regulatory environments.

\section*{Acknowledgements}
The authors are grateful to the Eldorado Research Institute.


\bibliography{sections/refs.bib}                     


\end{document}